\documentclass[pra,nofootinbib,showpacs,showkeys,draft]{revtex4}

\begin{document}
%
%
\title{Mixing angle between $^3P_1$ and $^1P_1$ in HQET}
%
%
\author{Takayuki Matsuki}
\email[E-mail: ]{matsuki@tokyo-kasei.ac.jp}
\affiliation{Tokyo Kasei University,
1-18-1 Kaga, Itabashi, Tokyo 173, JAPAN}
\author{Toshiyuki Morii}
\email[E-mail: ]{morii@kobe-u.ac.jp}
\affiliation{Graduate School of Human Development and Environment,
Kobe University,\\ Nada, Kobe 657-8501, JAPAN}
\author{Kohichi Seo}
\email{seo@gifu-cwc.ac.jp}
\affiliation{Gifu City Women's College, 7-1 Hito-ichiba Kitamachi,
Gifu 501-0192, JAPAN}

\date{\today}

\begin{abstract}
Some claim that there are two independent mixing angles 
($\theta = 35.3^\circ$, $-54.7^\circ$) between $^3P_1$ and $^1P_1$
states of heavy-light mesons in heavy quark symmetric limit, and others claim
there is only one ($\theta = 35.3^\circ$).
We clarify the difference between these two and suggest which should be
adopted. General arguments on the mixing angle between $^3L_L$ and $^1L_L$ of
heavy-light mesons are given in HQET and a general relation is derived in
heavy quark mass limit as well as that including the first order correction in
$1/m_Q$.
\end{abstract}

\preprint{}
\maketitle

%
\section{Introduction} \label{intro}
%

The $P$ states have a rich structure because combined with quark spins they
form four $P$ states, i.e., $^3P_0$, $^3P_1$, $^3P_2$, and $^1P_1$, and also
because there is an interesting feature of mixing between two $1^+$ states.
The explicit study on the mixing between $^3P_1$ and $^1P_1$ states in the context of a
heavy-light system is given by Rosner\cite{R86} and is restudied a few years later
in \cite{GK91} by taking into account more states, $D$, $D_s$, $B$, and $B_s$ mesons.
Here we illustrate their idea using notations of \cite{GK91} in which
as well as in \cite{R86}
they have assumed that dominant interaction between heavy and light quarks is
non-relativistic spin-orbit terms, which give mass for a heavy-light meson:
\begin{eqnarray}
 {H_{SO}} &=& \frac{4}
{3}\frac{{{\alpha _s}}}
{{{r^3}}}\frac{{\left( {{{\vec S}_q} + {{\vec S}_Q}} \right)\cdot\vec L}}
{{{m_q}{m_Q}}} + \frac{1}
{4}\left( {\frac{4}
{3}\frac{{{\alpha _s}}}
{{{r^3}}} - \frac{b}
{r}} \right)\left[ {\left( {\frac{1}
{{m_q^2}} + \frac{1}
{{m_Q^2}}} \right)\left( {{{\vec S}_q} + {{\vec S}_Q}} \right)\cdot\vec L + \left( {\frac{1}
{{m_q^2}} - \frac{1}
{{m_Q^2}}} \right)\left( {{{\vec S}_q} - {{\vec S}_Q}} \right)\cdot\vec L} \right]
\nonumber \\
 &=& H_{SO}^q \vec S_q\cdot\vec L + H_{SO}^Q \vec S_Q\cdot\vec L.
 \label{HSO}
\end{eqnarray}
Taking the heavy quark mass limit ($m_Q\to \infty$), we are left only with the first
term in the second line of Eq. (\ref{HSO}).
Assuming other interaction terms including kinetic terms give a constant $M_0$
contribution, then they give the
following relation between mass eigenstates and angular momentum
eigenfunctions.\cite{GK91}
\begin{eqnarray}
\left( {\begin{array}{*{20}{c}}
   {M\left( {^3{P_1}} \right)}  \\
   {M\left( {^1{P_1}} \right)}  \\

 \end{array} } \right) = \left( {\begin{array}{*{20}{c}}
   {{M_0} - \left\langle {H_{SO}^q} \right\rangle } & { - \sqrt 2 \left\langle {H_{SO}^q} \right\rangle }  \\
   { - \sqrt 2 \left\langle {H_{SO}^q} \right\rangle } & {{M_0}}  \\

 \end{array} } \right)\left( {\begin{array}{*{20}{c}}
   {^3{P_1}}  \\
   {^1{P_1}}  \\
 \end{array} } \right), \label{GKmass}
\end{eqnarray}
which is, as shown later, translated into
\begin{eqnarray}
\left( {\begin{array}{*{20}{c}}
   {\left| {{j^P=1^ + },j_\ell = {{\frac{1}
{2}} }} \right\rangle }  \\
   {\left| {{j^P=1^ + },j_\ell = {{\frac{3}
{2}} }} \right\rangle }  \\

 \end{array} } \right) = \left( {\begin{array}{*{20}{c}}
   {\cos \theta } & {\sin \theta }  \\
   { - \sin \theta } & {\cos \theta }  \\

 \end{array} } \right)\left( {\begin{array}{*{20}{c}}
   {\left| {^3{P_1}} \right\rangle }  \\
   {\left| {^1{P_1}} \right\rangle }  \\

 \end{array} } \right),
 \label{GKrel}
\end{eqnarray}
with two mixing angles,
\begin{equation}
  \theta = \arctan\left(\frac{1}{\sqrt{2}}\right)=35.3^\circ {\rm ~~or~~}
  \theta = \arctan\left(-\sqrt{2}\right)=-54.7^\circ, \label{mixGK}
\end{equation}
where the left hand side of Eq. (\ref{GKrel}) is the mass eigenstate and is specified in
terms of eigenvalues of a total anguluar momentum $\vec j$ of a $Q\bar q$ bound state,
$\vec j_\ell$ which stands for the light quark total angular momentum,
$\vec j_\ell=\vec L+\vec S_q$, whose square
is conserved in heavy quark symmetric limit, and the parity $P$.
Here $\vec S_q$ is a light quark spin.
The vector of the right hand side of Eq. (\ref{GKrel}) denoted as
$\left|^{2S+1}L_j \right> $ is specified in terms of
eigenvalues of a light quark angular momentum $\vec L$, a sum of intrinsic quark spins
$\vec S=\vec S_q+\vec S_Q$, and a total angular momentum $\vec j$ of the heavy-light meson.

On the other hand, using the heavy quark symmetry we have derived the
relation\footnote{Sign change of $\sin\theta$ in this equation
can be absorbed into state redefinitions so that the form of an orthogonal matrix
becomes the same as that of Eq. (\ref{GKrel}).} equivalent to
Eq. (\ref{GKrel}),\cite{MKM88,MM97}
\begin{eqnarray}
\left( {\begin{array}{*{20}{c}}
   {\left| {{j^P=1^ + },j_\ell = {{\frac{1}
{2}} }} \right\rangle }  \\
   {\left| {{j^P=1^ + },j_\ell = {{\frac{3}
{2}} }} \right\rangle }  \\

 \end{array} } \right) = \left( {\begin{array}{*{20}{c}}
   {\cos \theta } & { - \sin \theta }  \\
   {\sin \theta } & {\cos \theta }  \\

 \end{array} } \right)\left( {\begin{array}{*{20}{c}}
   {\left| {^3{P_1}} \right\rangle }  \\
   {\left| {^1{P_1}} \right\rangle }  \\

 \end{array} } \right) .
 \label{MMrel1}
\end{eqnarray}
but with only one mixing angle,
\begin{equation}
  \theta = \arctan\left(\frac{1}{\sqrt{2}}\right)=35.3^\circ . \label{mixMM}
\end{equation}
Equations (\ref{GKrel}) and (\ref{MMrel1}) are equivalent to each other but
Eq.~(\ref{mixMM}) is more restrictive than Eq.~(\ref{mixGK}).
We would like to solve the origin of this discrepancy and give a reasonable
interpretation which should be adopted for the heavy-light mesons.

%
\section{Mass Matrix by Rosner or Godfrey and Kokoski} \label{GK}
%

The expression of Eq. (\ref{GKmass}) is very confusing in the sense that 1)
there are no eigenstates on the l.h.s. of the equation, and 2) the eigenvalues
on the l.h.s., $M(^3P_1)$ and $M(^1P_1)$, are
written with explicit arguments $^3P_1$ and $^1P_1$.
They have assumed that the upper and lower components on the l.h.s. are
dominated by $^3P_1$ and $^1P_1$ from the beginning, respectively.

To better understand Eq. (\ref{GKmass}), we introduce ket vectors as eigenstates
with angular momentum quantum numbers, and an orthogonal
matrix, $U$, to diagonalize the mass matrix. 
We rewrite Eq. (\ref{GKmass}) as an eigenvalue equation in an operator form
so that everybody is on the same footing.
\begin{eqnarray}
  \left(M_0 + H^q_{SO} \vec L\cdot\vec S_q\right) \left|\psi\right\rangle = 
  \lambda \left|\psi\right\rangle, \quad
  \left|\psi\right\rangle = \alpha \left| {^3{P_1}} \right\rangle +
  \beta \left| {^1{P_1}} \right\rangle, \quad
  (\alpha^2+\beta^2=1 ), \label{eigen}
  \label{GKmass1}
\end{eqnarray}
where $\left|\psi\right\rangle$ is a wave function expanded in terms of
$\left|^{2S+1}L_j\right\rangle$, and $\alpha$ and $\beta$ are constant coefficients.
The mass Hamiltonian is defined by $2\times 2$ matrix of the r.h.s. of Eq. (\ref{GKmass}),
whose matrix elements are expectation values of
$M_0+H^q_{SO}\vec L\cdot\vec S_q$ between
$\left| {^3{P_1}} \right\rangle$ and $\left| {^1{P_1}} \right\rangle$.
Though it might be a rather redundant explanation shown below to solve Eq.~(\ref{eigen}), we believe that it clarifies the reason why two mixing angles appear.

Now we can reexpress Eq. (\ref{GKmass1}) in the following eigenvalue equation in which
all quantities are constant.
\begin{eqnarray}
  && M P = \lambda P {\rm ~~or~~} M_D P' = \lambda  P', \label{EigenGK}
\end{eqnarray}
where
\begin{eqnarray}
  &&  M_D\equiv UMU^T, \quad P'=UP, \quad
  M = \left( {\begin{array}{*{20}{c}}
   {{M_0} - \left\langle {H_{SO}^q} \right\rangle } & { - \sqrt 2 \left\langle {H_{SO}^q} \right\rangle }  \\
   { - \sqrt 2 \left\langle {H_{SO}^q} \right\rangle } & {{M_0}}  \\
 \end{array} } \right), \nonumber \\
  && U = \left( {\begin{array}{*{20}{c}}
   {\cos \theta } & {\sin \theta }  \\
   { - \sin \theta } & {\cos \theta }  \\

 \end{array} } \right),\;P = \left( {\begin{array}{*{20}{c}}
   \alpha  \\
   \beta  \\

 \end{array} } \right),\;P' = UP = \left( {\begin{array}{*{20}{c}}
   \alpha'  \\
   \beta'  \\

 \end{array} } \right), \quad (\alpha'^2+\beta'^2=1 )
  \label{MixGK}
\end{eqnarray}
When one solves an eigenvalue equation $MP=\lambda P$, we obtain,
\begin{eqnarray}
&& \lambda = {M_0} - 2\left\langle {H_{SO}^q} \right\rangle
{\rm ~~or~~}\;{M_0} + \left\langle {H_{SO}^q} \right\rangle ;
\;P' = \left( {\begin{array}{*{20}{c}}
   1  \\
   0  \\
 \end{array} } \right)
{\rm ~~or~~}\;P' = \left( {\begin{array}{*{20}{c}}
   0  \\
   1  \\

 \end{array} } \right), \label{solM1} \nonumber \\
%
&& P = \frac{1}{{\sqrt 3 }}\left( {\begin{array}{*{20}{c}}
   {\sqrt 2 }  \\
   1  \\
\end{array}} \right)\quad {\rm{or}}\quad P = \frac{1}{{\sqrt 3 }}\left( {\begin{array}{*{20}{c}}
   { - 1}  \\
   {\sqrt 2 }  \\
\end{array}} \right), \label{eigenP}
\end{eqnarray}
respectively, and we have only one mixing angle,
\begin{eqnarray}
  \theta=\arctan\left(1/\sqrt{2}\right)=35.3^\circ.
\end{eqnarray}
Inserting Eq. (\ref{eigenP}) into Eq. (\ref{eigen}), we have eigenfunctions
$\left|\psi\right\rangle=\left|^{2S+1}P_1\right\rangle'$ as,
\begin{eqnarray}
\left| {^3{P_1}} \right\rangle ' \equiv \frac{1}
{{\sqrt 3 }}\left( {\sqrt 2 \left| {^3{P_1}} \right\rangle  + \left| {^1{P_1}} \right\rangle } \right),
\quad
\left| {^1{P_1}} \right\rangle ' \equiv \frac{1}
{{\sqrt 3 }}\left( { - \left| {^3{P_1}} \right\rangle  + \sqrt 2 \left| {^1{P_1}} \right\rangle } \right), \label{solM2}
\end{eqnarray}
where we have named eigenstates $\left|^{2S+1}P_1\right\rangle'$ on the l.h.s. of
Eq.~(\ref{solM2}) according to which coefficient of eigenstates 
$\left|^{2S+1}P_1\right\rangle$ on the r.h.s. is larger, e.g., in the r.h.s. of the
first equation
a coefficient of $\left|^3P_1\right\rangle$ $\left(\sqrt{2/3}\right)$ is lager than
that of $\left|^1P_1\right\rangle$ $\left(\sqrt{1/3}\right)$, thus we call this
$\left|^3P_1\right\rangle'$.

Another way to solve Eq. (\ref{EigenGK}) $M_DP'=\lambda P'$ is to require vanishing
off-diagonal elements of $M_D$, which gives the following two mixing angles $\theta$
as in \cite{R86,GK91},
\begin{eqnarray}
  \theta_1 = \arctan\left(\frac{1}{\sqrt{2}}\right)=35.3^\circ {\rm ~~or~~}
  \theta_2 = \arctan\left(-\sqrt{2}\right)=-54.7^\circ.
\end{eqnarray}
In order to check whether they are independent or not, we may rewrite the eigenvalue
euqation $M_DP'=\lambda P'$ for each angle, which are given as follows.
In the case of $\tan \theta  =\tan \theta_1  = \frac{1}{{\sqrt 2 }}$, the diagonalized mass matrix
and eigenvectors are given by,
\begin{eqnarray}
 M_{D1} &=& U_1MU_1^T = \left( {\begin{array}{*{20}{c}}
   {{M_0} - 2\left\langle {H_{SO}^q} \right\rangle } & 0  \\
   0 & {{M_0} + \left\langle {H_{SO}^q} \right\rangle }  \\
 \end{array} } \right), \quad
   P_1' = 
 \left( {\begin{array}{*{20}{c}}
   1  \\
   0  \\
 \end{array} } \right) {\rm ~~or~~}
 \left( {\begin{array}{*{20}{c}}
   0  \\
   1  \\
 \end{array} } \right),
\nonumber \\
  P_1 &=& U_1^T P_1'=\frac{1}{\sqrt{3}} \left( {\begin{array}{*{20}{c}}
   \sqrt{2}  \\
   1  \\
 \end{array} } \right) {\rm ~~or~~}
  \frac{1}{\sqrt{3}} \left( {\begin{array}{*{20}{c}}
   -1  \\
   \sqrt{2}  \\
 \end{array} } \right) ,
 \label{eigen1}
\end{eqnarray}
respectively, with $U_1=U(\theta=\theta_1)$ in Eq. (\ref{MixGK}).
In the case of $\tan \theta  =\tan \theta_2  = -\sqrt{2}$,
those are given by,
\begin{eqnarray}
 M_{D2} &=& U_2MU_2^T = \left( {\begin{array}{*{20}{c}}
   {{M_0} + \left\langle {H_{SO}^q} \right\rangle } & 0  \\
   0 & {{M_0} - 2\left\langle {H_{SO}^q} \right\rangle }  \\
 \end{array} } \right), \quad
   P_2' = 
 \left( {\begin{array}{*{20}{c}}
   -1  \\
   0  \\
 \end{array} } \right) {\rm ~~or~~}
 \left( {\begin{array}{*{20}{c}}
   0  \\
   1  \\
 \end{array} } \right),
\nonumber \\
  P_2 &=& U_2^T P_2' = \frac{1}{\sqrt{3}} \left( {\begin{array}{*{20}{c}}
   -1  \\
   \sqrt{2}
 \end{array} } \right) {\rm ~~or~~}
  \frac{1}{\sqrt{3}} \left( {\begin{array}{*{20}{c}}
   \sqrt{2}  \\
   1  \\
 \end{array} } \right) , \label{eigen2}
\end{eqnarray}
respectively, with $U_2=U(\theta=\theta_2)$ in Eq. (\ref{MixGK}).
Multiplying the following matrix $U_0$ on $M_{D2}$, $P'_2$, and $U_2$
in Eq. (\ref{eigen2}) as,
\begin{eqnarray}
{U_0} = 
\left( {\begin{array}{*{20}{c}}
   0 & 1  \\
   { - 1} & 0  \\
\end{array}} \right) =
\left( {\begin{array}{*{20}{c}}
   \cos 90^\circ & \sin 90^\circ  \\
   { - \sin 90^\circ} & \cos 90^\circ  \\
\end{array}} \right) , \quad
  M_{D1}={U_0}{M_{D2}}U_0^T,\quad P_1'={U_0}P_2' , 
  \quad U_1 = U_0 U_2, \label{transf2}
\end{eqnarray}
we can reproduce Eq. (\ref{eigen1}). Hence Eqs.~(\ref{eigen1}) and (\ref{eigen2}) are
equivalent to each other, which means that two mixing angles are also equivalent.
Actually $\theta_2=\theta_1-90^\circ $ as easily seen from Eq.~(\ref{transf2}).
This is consistent with the solution given by
Eqs. (\ref{solM1}) $\sim$ (\ref{solM2}) with the mixing angle $\tan\theta=1/\sqrt{2}$
when solving the eigenvalue equation $MP=\lambda P$.

When one tries to identify which eigenstate corresponds to a lower-mass or
higher-mass state as in Refs. \cite{R86,GK91},
it does not matter which angle one adopts. It depends only on sign of
$\left<H_{SO}^q\right>$.
By looking at
Eqs.~(\ref{eigen1}, \ref{eigen2}), one finds that if $\left<H_{SO}^q\right> > 0$,
then the lower-mass state is identified as $\left|^3P_1\right>'$ and
the higher-mass as $\left|^1P_1\right>'$.
On the other hand, if $\left<H_{SO}^q\right> < 0$, the lower-mass
state is identified as $\left|^1P_1\right>'$ and the higher as
$\left|^3P_1\right>'$ irrespective of a mixing angle.

There is a way to determine which state
($\left| {^3{P_1}} \right\rangle '$ or $\left| {^1{P_1}} \right\rangle '$) corresponds
to which heavy quark symmetric state ($j_\ell^P=(1/2)^+$ or $(3/2)^+$).
This is done by expanding heavy quark symmetric states $\left|j,j_\ell, j_z\right>$
in terms of states $\left|j,S,j_z \right>$ with $\vec S=\vec S_q+\vec S_Q$, i.e.,
by calculating the $6-j$ symbols,
which is given in Appendix of \cite{CJ03} as,
\begin{eqnarray}
\left( {\begin{array}{*{20}{c}}
   {\left| {j = L ,{j_\ell } = L  - 1/2,m} \right\rangle }  \\ 
   {\left| {j = L ,{j_\ell } = L  + 1/2,m} \right\rangle }  \\ 
\end{array} } \right) = \frac{1}{{\sqrt {2j + 1} }}\left( {\begin{array}{*{20}{c}}
   {\sqrt {L + 1} } & {\sqrt L }  \\ 
   { - \sqrt L } & {\sqrt {L + 1} }  \\ 
\end{array} } \right)\left( {\begin{array}{*{20}{c}}
   {\left| {j = L ,S = 0,m} \right\rangle }  \\ 
   {\left| {j = L ,S = 1,m} \right\rangle }  \\ 
\end{array} } \right) .
\end{eqnarray}
By substituting $L=1$, we obtain Eq.~(\ref{GKrel}).
Therefore even discussions given by \cite{R86,GK91} are enough to uniquely determine
the relation between heavy quark symmetric states $\left| {{j^P},j_\ell} \right\rangle$
and non-relativistic states $\left|^{2S+1}L_j\right\rangle$ in heavy quark symmetric limit, which is given by Eq. (\ref{GKrel}) with only one mixing angle
Eq.~(\ref{mixMM}).\cite{XL09}

%
\section{Mixing between $^3L_L$ and $^1L_L$ in HQET} \label{MM}
%

In the relativistic potential model studied by us more than ten years ago, we have
derived the relativistic equation for a $Q\bar q$ bound state in the heavy quark
symmetric limit ($m_Q\to \infty$) treating a light quark as relativistic and
a heavy quark as static.\cite{MM97} In that equation {\it the angular component
is completely solved} and is given by the eigenfunction $y_{jm}^k$.
Because a heavy quark is treated as static in heavy quark limit, a bound state
wave function can be separated into (heavy-quark) energy positive and negative
components and the lowest non-trivial order wave function is naturally given by
a positive energy component which has $2\times 4$ spinor components. In order to
classify the states in
terms of a non-relativistic $^{2S+1}L_j$ state, only the upper $2\times 2$ component of
the wave function is necessary. The relation between $y_{jm}^k$ and angular momentum
eigenfunctions is uniquely determiend to be,
\begin{equation}
  \left( {\matrix{{y_{j\,m}^{-(j+1)}}\cr {y_{j\,m}^j}\cr}} \right)
  =U\left( {\matrix{{Y_j^m}\cr 
  {\vec \sigma \cdot \vec Y_{j\,m}^{(\rm M)}}\cr}} \right),
  \quad\left( {\matrix{{y_{j\,m}^{j+1}}\cr {y_{j\,m}^{-j}}\cr}} \right)
  =U\left( {\matrix{{\vec \sigma \cdot \vec Y_{1\,m}^{(\rm L)}}\cr
  {\vec \sigma \cdot \vec Y_{j\,m}^{(\rm E)}}\cr}} \right), \quad
  U = \frac{1}
  {{\sqrt {2j+1} }}\left( {\begin{array}{*{20}{c}}
   {\sqrt {j+1} } & {\sqrt{j}}  \\
   { -\sqrt{j}} & {\sqrt {j+1} }  \\
 \end{array} } \right). \label{MMSrel}
\end{equation}
That is, this is the definition of the eigenfunction $y_{jm}^k$. When $j=1$, we have
the following relation between the eigenstates (l.h.s.)
respecting the heavy quark symmetry and the non-relativistic states (r.h.s.)
described in terms of $^3P_1$ and $^1P_1$.
\begin{equation}
  \left( {\matrix{{y_{1\,m}^{-2}}\cr {y_{1\,m}^1}\cr}} \right)
  =U\left( {\matrix{{Y_1^m}\cr 
  {\vec \sigma \cdot \vec Y_{1\,m}^{(\rm M)}}\cr}} \right)
  \qquad {\rm with~~}
  U = \frac{1}
  {{\sqrt 3 }}\left( {\begin{array}{*{20}{c}}
   {\sqrt 2 } & {1}  \\
   { -1} & {\sqrt 2 }  \\
 \end{array} } \right) . \label{MMSrel2}
\end{equation}
Here $Y^{(E),\;(M),\;(L)}$ are spinor representations of an intrinsic spin $s=1$
particle with a total angular momentum $j$, i.e., photon's
wave function with a total angular momentum $j$.
Here $Y^{(E),\;(M),\;(L)}$ have parities, $(-)^{j+1}$, $(-)^{j}$, and $(-)^{j+1}$,
respectively, and $Y_j^m$ has parity,  $(-)^{j}$, i.e., the same as $Y^{(M)}$.
That is, $Y^{(M)}$ is a spinor representation of $^3P_1$ and so is
$1_{2\times 2}\times Y_1^m$ that of
$^1P_1$, while wave functions on the l.h.s., $y_{1\,m}^{-2}$ and $y_{1\,m}^{1}$
correspond to $j_\ell=3/2$ and $1/2$ with $j^P=1^+$, respectively.
Here we have used the relation between $j_\ell$ and $k$,\cite{MMMS04}
\begin{equation}
  j_\ell = \left| k \right| - \frac{1}{2}. \label{k-jrel}
\end{equation}
Equation (\ref{MMSrel2}) is our result which is equivalent to 
Eq. (\ref{MMrel1}). The mixing angle is given by
$\theta=\arctan\left(1/\sqrt{2}\right)=35.3^\circ$ 
that is not a "magic number" as called in Refs. \cite{BBP03,CS05}, which
is derived from the relation between eigenstates with a $k$ quantum number and
$^{2S+1}L_j$ states.

Using the first equation of Eq.~(\ref{MMSrel}), we can write down a general relation
between heavy quark symmetric states and non-relativistic states $^3L_L$ and
$^1L_L$ as,
\begin{equation}
\left( {\begin{array}{*{20}{c}}
   {\left| {y_{L{\kern 1pt} m}^L} \right\rangle }  \\ 
   {\left|  {y_{L{\kern 1pt} m}^{ - (L + 1)}} \right\rangle }  \\ 
\end{array} } \right) = \left( {\begin{array}{*{20}{c}}
   {\left| {{L^P},{j_\ell } = L - \frac{1}{2}} \right> }  \\ 
   {\left| {{L^P},{j_\ell } = L + \frac{1}{2}} \right> }  \\ 
\end{array} } \right) = \frac{1}{{\sqrt {2L + 1} }}\left( {\begin{array}{*{20}{c}}
   {\sqrt {L + 1} } & { - \sqrt L }  \\ 
   {\sqrt L } & {\sqrt {L + 1} }  \\ 
\end{array} } \right)\left( {\begin{array}{*{20}{c}}
   {\left| {^3{L_L}} \right\rangle }  \\ 
   {\left| {^1{L_L}} \right\rangle }  \\ 
\end{array} } \right) , \quad P= (-1)^{L + 1} ,
\end{equation}
which gives Eq.~(\ref{MMrel1}) when $j=L=1$.
Here we have used $P= (-1)^{|k| + 1}\;{k}/{|k|}$ with $k=L$.\cite{MMMS04}

In our model\cite{MM97}, a spin doublet $(0^+, 1^+)$ degenerates and so does another
spin double $(1^+, 2^+)$ in heavy quark symmetric limit, which are corresponding to
$j_\ell^P=(1/2)^+$ and $(3/2)^+$ multiplets, respectively. Our most recent numerical
calculations\cite{MMS07} show that $M\left((1/2)^+\right) < M\left((3/2)^+\right)$
in the cases of $c\bar q$ and $b\bar q$ which is equivalent
to $M\left(\left|{^3{P_1}} \right>'\right) < M\left(\left|{^1{P_1}} \right>'\right)$.
These values of $M$ are degenerate eigenvalues of a first-order differential equation
and can not be predicted beforehand by just looking at the equation.

There appear a couple of quantum numbers to distinguish
heavy-light mesons, which is summarized in TABLE \ref{table1}. Here $j$ stands for
a total angular momentum, $P$ its parity, $k$ a quantum number whose relation with other
quantum numbers is given by, e.g., Eq. (\ref{k-jrel}),\cite{MMMS04,JJ} $j_\ell^P$ a total
angular momentum of a light quark with parity $P$, and $^{2S+1}L_j$ a non-relativistic
quantum number describing a total intrinsic spin $S$, an internal angular momentum $L$,
and a total angular momentum $j$. 

\begin{table*}[t]
\caption{States classified by various quantum numbers.}
\label{table1}
\begin{tabular*}{13cm}{c|@{\extracolsep{\fill}}cccccccccc}
\hline
\hline
  \makebox[1.7cm]
 {$j^P$}   &   $0^-$   &   $1^-$   &   $0^+$   &   
  $1^+$   &   $1^+$   &   $2^+$   &   $1^-$   &   $2^-$ &  $2^-$  &  $3^-$   \\
  $k$   &   -1   &   -1   &   1   &   
   1    &   -2   &   -2   &   2   &   2 &   -3   &   -3   \\
  $j_\ell^{P}$ & ${1\over2}^-$ & ${1\over2}^-$ & ${1\over2}^+$ & 
  ${1\over2}^+$ & ${3\over2}^+$ & ${3\over2}^+$ & ${3\over2}^-$ & ${3\over2}^-$
  & ${5\over2}^-$ & ${5\over2}^-$ \\ \hline
  $^{2S+1}L_j$ & $^1S_0$ & $^3S_1$  & $^3P_0$  & $\left(^3P_1, ~^1P_1\right)$ &
   $\left(^1P_1, ~^3P_1\right)$  & $^3P_2$  & $^3D_1$  & $\left(^3D_2, ~^1D_2\right)$
  & $\left(^1D_2, ~^3D_2\right)$  & $^3D_3$ \\ \hline 
\hline
\end{tabular*}
\end{table*}

%
\section{Breaking of Heavy Quark Symmetry} \label{BHQS}
%

Let us briefly discuss what the mixing angle tells us when the heavy quark symmetry
is broken.
A general mixing angle between $^3L_L$ and $^1L_L$ in HQET is given by
$\tan\theta=\sqrt{L/(L+1)}$ as readily seen from Eq. (\ref{MMSrel}) and when one
takes into account breaking of the heavy quark symmetry, it is given by
\begin{eqnarray}
  \tan \left( {{\theta _1} + \delta \theta } \right) = \sqrt{\frac{ L }
{L + 1} } + \frac{{\left( {2L + 1} \right)}}
{{L + 1}} \delta \theta ,\quad \tan {\theta _1} = \sqrt{ \frac{{ L }}
{{ {L + 1} }}}, \quad
  \delta \theta = O\left(\frac{1}{m_Q} \right) .
\end{eqnarray}
Because $\tan\theta_1=\sqrt{L/(L+1)}$ is the result of heavy quark symmetry,
$\delta\theta$ gives mixing between states with different $j_\ell$ as
\begin{eqnarray}
\left( {\begin{array}{*{20}{c}}
   {\left| {{L^ P },{j_\ell } = L - \frac{1}{2}} \right\rangle '}  \\
   {\left| {{L^ P },{j_\ell } = L + \frac{1}{2}} \right\rangle '}  \\
\end{array}} \right) = \left( {\begin{array}{*{20}{c}}
   1 & { -  \delta \theta }  \\
   { \delta \theta } & 1  \\
\end{array}} \right)\left( {\begin{array}{*{20}{c}}
   {\left| {{L^ P },{j_\ell } = L - \frac{1}{2}} \right\rangle }  \\
   {\left| {{L^ P },{j_\ell } = L + \frac{1}{2}} \right\rangle }  \\
\end{array}} \right),   \quad P=  (-1)^{L + 1} ,
\end{eqnarray}
where $k=j=L$ is assumed. See \cite{BELLE07} for discussions on this kind of mixing.

%
\section{Summary} \label{summary}
%

We conclude from the previous sections' results that the heavy quark symmetry
can uniquely determine the relation between heavy quark symmetric eigenstates and
states with $^{2S+1}P_1$ with the mixing angle
$\theta=35.3^\circ$
between $^3P_1$ and $^1P_1$ as shown by Eq. (\ref{MMrel1}) as,
\begin{eqnarray*}
\left( {\begin{array}{*{20}{c}}
   {\left| {{1^ + },j_\ell  = {{\frac{1}
{2}} }} \right\rangle }  \\
   {\left| {{1^ + },j_\ell  = {{\frac{3}
{2}} }} \right\rangle }  \\

 \end{array} } \right) = \left( {\begin{array}{*{20}{c}}
   {\cos \theta } & { - \sin \theta }  \\
   {\sin \theta } & {\cos \theta }  \\

 \end{array} } \right)\left( {\begin{array}{*{20}{c}}
   {\left| {^3{P_1}} \right\rangle }  \\
   {\left| {^1{P_1}} \right\rangle }  \\

 \end{array} } \right) 
 \quad {\rm with~~} \tan\theta = \frac{1}{\sqrt{2}}.
\end{eqnarray*}

In heavy quark symmetric limit, our relativistic potential model\cite{MMS07}
predicts the lower-mass state is $\left| {^3{P_1}} \right> '$ and the higher-mass
state $\left| {^1{P_1}} \right\rangle '$, while the model with the Breit-Fermi type
non-relativistic potential model\cite{R86,GK91} predicts either
$\left| {^3{P_1}} \right\rangle '$ or $\left| {^1{P_1}} \right\rangle '$ as the
lower mass state depending on sign of $\left<H_{SO}^q\right>$.

Finally let us clarify the reason why the mass matrix given by Rosner\cite{R86} or
Godfrey and
Kokoski\cite{GK91} gives the same eigenstates $\left| {^3{P_1}} \right\rangle '$
and $\left| {^1{P_1}} \right\rangle '$ as our model.
Interactions including only the spin-orbit terms can be diagonalized by $y_{jm}^k$
because these are eigenfunctions of the operator $\vec L\cdot\vec \sigma_q$ as,
\begin{eqnarray}
  \vec L\cdot\vec\sigma_q \left| y_{jm}^k \right\rangle = 
  -(k+1) \left| y_{jm}^k \right\rangle .
\end{eqnarray}
%
Because
\begin{eqnarray*}
  \left| y_{1m}^{-2} \right> \equiv \left| {^1{P_1}} \right\rangle ', \quad
  \left| y_{1m}^{1} \right> \equiv \left| {^3{P_1}} \right\rangle ' ,
\end{eqnarray*}
the operator $\vec L\cdot\vec\sigma_q$ has the following eigenvalues,
\begin{eqnarray*}
  \vec L\cdot\vec\sigma_q \left| {^3{P_1}} \right\rangle ' = 
  -2 \left| {^3{P_1}} \right\rangle ', \quad
  \vec L\cdot\vec\sigma_q \left| {^1{P_1}} \right\rangle ' =
   \left| {^1{P_1}} \right\rangle ' .
\end{eqnarray*}
Hence we have
\begin{eqnarray}
  \vec j_\ell~^2 = \left(\vec L+\vec S_q\right)^2=\frac{3}{4}, \quad \frac{15}{4},
  {\rm ~~or~~} j_\ell = \frac{1}{2}, \quad \frac{3}{2},
\end{eqnarray}
respectively. Here $L=1$ and $\vec S_q=\vec\sigma_q/2$.
This simply is the reason why they have obtained the same eigenstates
$\left| {^3{P_1}} \right\rangle '$ and $\left| {^1{P_1}} \right\rangle '$ as our model.
The functions $y_{jm}^k$ specified by $k$,\cite{MMMS04,JJ} $j$, and $m$ quantum numbers
are equivalent to the heavy quark eigenstates specified by $j^P$ and $j_\ell$ as,
\begin{eqnarray}
  \left| y_{jm}^k \right>=\left| {{j^P},j_\ell} \right\rangle \quad \left
  ({\rm ~with~~} j = \left| k \right|
  \;{\text{or}}\;\left| k \right| - 1,\quad {j_\ell } =
  \left| k \right| - \frac{1}{2},\quad 
  P = \frac{k}{|k|} (-1)^{|k| + 1} \right)  ,
\end{eqnarray}
where $k\ne 0$ and we have omitted a quantum number $m$ on the r.h.s..

\acknowledgments 
One of the authors (T. Matsuki) would like to thank Dr. Xiang Liu for
discussions.

\def\Journal#1#2#3#4{{#1} {\bf #2}, #3 (#4)}
\def\NIM{Nucl. Instrum. Methods}
\def\NIMA{Nucl. Instrum. Methods A}
\def\NPB{Nucl. Phys. B}
\def\PLB{Phys. Lett. B}
\def\PRL{Phys. Rev. Lett.}
\def\PRD{Phys. Rev. D}
\def\PTP{Prog. Theor. Phys.}
\def\ZPC{Z. Phys. C}
\def\EPJ{Eur. Phys. J. C}
\def\PR{Phys. Rept.}
\def\IJM{Int. J. Mod. Phys. A}

\end{document}